\documentclass[a4paper,twocolumn,10pt]{article}
\usepackage[authoryear]{natbib}
\usepackage{authblk}
\usepackage{microtype}
\usepackage{geometry}
\usepackage[colorlinks=true, linkcolor=black, urlcolor=black, citecolor=black]{hyperref}
\usepackage{url}
\usepackage{graphicx}
\usepackage{caption}
\usepackage{subcaption}

\begin{document}

\title{Building AI-based advisory services for smallholder farmers: Technical learnings  from the AIEP Initiative}

\author[ ]{AIEP Initiative\thanks{The AIEP initiative was a collaborative undertaking and this paper summarises inputs and thoughts of many who were involved but especially the members of the cohorts. This is to represent their shared contribution. The remaining authors are the GIZ core team with support from CLEAR Global and the team at the Gates Foundation and are listed alphabetically.}}
\author[1]{Stewart Collis}
\author[2]{Florence Kinyua}
\author[2]{Vikram Kumar}
\author[1]{Howard Lakougna}
\author[2]{Christian Merz}
\author[2]{Kirti Pandey\thanks{Current affiliation: Independent}}
\author[3]{Christian Resch}
\affil[1]{Gates Foundation}
\affil[2]{Deutsche Gesellschaft für International Zusammenarbeit}
\affil[3]{CLEAR Global}


\maketitle

\begin{abstract}

We report technical learnings from five AI-based agricultural advisory MVPs deployed in Kenya and Bihar, India, under the AIEP Initiative. A ~800-farmer study found high user satisfaction (NPS $\sim$60). All solutions implement a modular two-part architecture: (i) an interface component (IVR / WhatsApp / app) with ASR→MT→TTS for multilingual voice access; and (ii) a reasoning component combining LLMs capabilities with query orchestration, external data (weather/soil/markets), and RAG over curated agricultural corpora. We describe key challenges: (a) latency, especially for voice; reductions were achieved via in-country hosting and audio minimization, but consistent <5s remains challenging; (b) language coverage: low-resource ASR/MT integration and nonstandard scripts hinder end-to-end quality; and (c) corpus curation: access, validation, and maintenance are labor-intensive, as well as provide recommendations on how to develop similar systems. We discuss common enablers including (a) data sharing, (b) common corpora, (c) better language AI and (d) evaluation and benchmarking. We also present golden Q\&A sets to evaluate LLM capabilities for smallholder agriculture.

\end{abstract}

\section{Introduction: What is this about?}
Most of the more than 570 million farms in the world are less than 2 hectares and family run \citep{Lowder2016WorldDev}, yet these smallholder farmers face diverse challenges, many of which have uncertainty and lack of information in common: What is the right strategy to contain an unknown disease? Which and how to use new seed varieties? How to behave in the face of changing weather patterns that defy traditional knowledge shaped by centuries-old reliable patterns?

There are often answers to these questions, but they often lie beyond the reach of many smallholder farmers because traditional agricultural extension systems struggle to provide their service and support at scale, with ratios of extension agents to farmers below 1 to 750 in India \citep{Nandi2019AgExtensionIndia} and 1 to 1000 in Kenya \citep{Kenya2012NASEP} with large variations by region.

Various digital solutions exist to close this gap, but many of those solutions struggle to provide the relevant, timely, and personalized information needed by smallholder farmers. At the same time, the effectiveness of video-mediated extension services like Digital Green’s have proven impacts on the uptake of knowledge and technology \citep{Abate2023WorldDev, VanCampenhout2021AJAE}. SMS have been shown to have a modest impact at low cost \citep{Fabregas2025DigitalInformation}.

The Agricultural Information Exchange Platform (AIEP) Initiative, funded by the Gates Foundation and implemented by GIZ’s 'FAIR Forward – Artificial Intelligence for All' Program with support from CLEAR Global, explored the potential of emerging artificial intelligence (AI) solutions to address this challenge and increased collaboration and exchange between partners working on these solutions to provide better digital services and specifically better agricultural advisory services.

As part of the AIEP initiative, five cohorts made up of organizations with different skills ranging from technology and agriculture to product management and user research have developed minimum viable products (MVPs) of AI-based agricultural advisory systems that take advantage of the most recent technological innovation and available agricultural content in Kenya and the State of Bihar, India. Those MVPs aimed to improve existing digital solutions by providing easier access through voice and natural language interfaces, allowing for two-way communication and more relevant, timely, and personalized information, ultimately contributing to supporting smallholder farmers in overcoming the challenges to improve their yields and livelihoods.

A first user study conducted by 60 decibels of over 800 smallholder farmers, 38\% of which were women, provided first promising evidence on the potential of AI-based advisory. The net promoter score (NPS)\footnote{The net promoter score is calculated based on response to the question “How likely would you be to recommend [...]?” on a scale from 0 to 10. The score is the share of 'detractors' (responses of 6 and less) of the share of 'promoters' (responses of 9 and greater; see \citep{Reichheld2003OneNumberYouNeedToGrow} for details)} as satisfaction measurement is 60 on average in cohorts which exceeds global (NPS 46), Kenyan (NPS 26) and Bihar (NPS 40) benchmarks of digital farmers services collected by 60 decibels. Importantly, women report a higher NPS than men. Qualitative interviews indicated that farmers often share the information received from AIEP solutions with their peers and encouraged them to also use the solution. 

Among the areas for improvement, the completeness and relevance of the information provided stands out. It is the main reason why farmers would not recommend solutions to others. In addition, 50\% of those who reported that they had not applied the information provided reported that the availability of inputs or financial constraints kept them from taking action. This indicates that the information provided is not always sufficiently tailored to the context and that AI-based advisory in addition to information will need integration with other services that alleviate constraints. Unsurprising yet concerning is as well that farmers with lower educational levels reported a higher likelihood that information was not easy to understand (overall 33\% of farmers), pointing to more work required to fulfill our original goals of providing solutions working for all smallholder farmers.

This study provides first encouraging data points, but will need to be complemented by further studies in the future. Like most studies, this also showed some potential limitations. For example, educational attainment is higher than the average smallholder farmer, with 30\% of the survey participants reporting a university, polytechnical, or other higher degree.

Despite limitations, the 60db study suggests that AI-based advisory has strong potential. Farmers seem satisfied with the solutions, stressing their availability and quality of information. The next step which various partners of the AIEP initiative are preparing now is rigorous evaluations of the impact on yields and livelihood to show the material benefit of the solutions.

This paper is one of two in which we aggregate insights and learnings that we and our partners have gathered in the last two years of building the MVPs. It aims to elucidate the technical foundations of AI-based agricultural advisory systems, drawing on the experiences and lessons learned from the AIEP project, while its companion paper describes and presents learnings about our innovation methodology (see \citet{AIEP_method_2025}). This paper explores the design and implementation of AI technologies in agriculture, discusses the challenges encountered, and provides recommendations for developing similar systems in other sectors. By sharing these insights, we hope to contribute to the broader discourse on leveraging AI for sustainable development.

\section{Background on the AIEP initiative}

The AIEP initiative started in 2023 by selecting four cohorts out of more than 100 applications received in response to a call for proposals. Opportunity International, SafariCom Digifarm and Gooey.ai later joined as the fifth cohort.
The key criteria for the selection of the eventual AIEP cohorts were technological innovation; end user and community orientation; gender-inclusivity; impact orientation; evidence; and user feedback; local representation; and long-term sustainability and clear ownership of the MVP.
The collective MVP development started in late 2023 and continued until early 2025. During this phase, we met monthly for shared print reviews and demos to exchange learnings and insights, and for a total of three gatherings to dedicate time for in-depth exchange and learning.

\section{Solutions descriptions: What are the 5 MVPs?}

AIEP consisted of five cohorts, four selected through the initial process, and the cohort consisting of Opportunity International, Safaricom Digifarm, and \href{https://gooey.ai}{Gooey.ai}, which joined the initiative later. 

\begin{itemize}
    \item dynAg, a partnership between IRRI, CIMMYT, Dexian India Technologies Private Ltd., Gramhal, IFFCO Kisan, H3i Technologies Private Limited, MICROX Foundation and Sumarth - The dynAG cohort focuses on Bihar only and is led by the CGIAR, namely the International Rice and Research Institute (IRRI) and CIMMYT. Dexian oversees application development (system integrator) supported by H3i Technologies (UX and LLM integration). End user engagement and farmer facing field support is covered through IFFCO Kissan and Sumarth utilizing ongoing partnerships, for example, with JEEVIKA. The dynAg platform is an extended version of the ‘matar’ application developed by H3i Technologies and is available via the ai.sakhi app and IVR.
    \item FarmerChat, a partnership between Digital Green, Karya, Gooey.ai, Gramvaani - The cohort included Digital Green, one of the leading non-profit organizations in the digital agriculture space along with its partners Karya Inc, Gooey.ai and Gramvaani. Gooey.ai brought deep expertise in LLM, tools and models including benchmarking and evaluation. The company is working with Karya Inc. for time-efficient fine-tuning and some parts of the ASR pipeline. Graamvani supported by developing an IVR pilot for the solution. The MVP is now available as the generative AI-powered chatbot FarmerChat. For further details, see \citet{Singh2024FarmerChat}.
    \item Tech4Her, a partnership between DeHaat and Dalberg - The cohort comprises two entities: DeHaat, a well-known agri-tech startup in India with a presence across various geographies, and Dalberg, which has developed a Human-Centered Design (HCD) toolkit for user research. The Minimum Viable Product (MVP) is an open-source architecture enabling customized, inclusive, and interactive information exchange by last mile women farmers and extension agents, interoperable with omni-channel interfaces. The cohort leveraged the extensive end-user knowledge of Dalberg and access to DeHaat's internal agricultural data.
    \item Viamo, Sahay, HarvestPlus, and Producers Direct - This cohort covered both pilot regions (Kenya and Bihar) and developed the MVP on top of Viamo’s IVR service. Sahaj contributed their tech expertise and played key roles in supporting the development of the speech technology and Large Language Model (LLM)-based RAG pipelines. HarvestPlus and Producers Direct implemented user testing with farmers through their network and contributed agricultural data sets. 
    \item Opportunity International, SafariCom DigiFarm and Gooey.ai - The cohort developed a WhatsApp based bot in Kenya that emerged out of a custom built solution tested in Malawi with 150 users. Opportunity International is an international NGO that has the role of integrator and subject matter expert on extension services and the deployment of AI bots for agricultural extension. DigiFarm, the Kenyan entity and wholly owned subsidiary of Safaricom, has local knowledge in Kenya and sources advisory content. In addition, DigiFarm has exceptional access to farmer communities and coordinates field outreach and research. Gooey.ai provides a low-code AI orchestration platform comprising technical services for the conversational interfaces, hosts RAG tools, function interfaces for external services and the WhatsApp interface, logs and analysis results, orchestrates knowledge, prompts, and natural language processing.
\end{itemize}

\section{Related work: What other solutions are there?}

Beyond the AIEP cohorts, there are various other partners that work on similar solutions to support smallholder farmers, including:
\begin{itemize}
    \item iSDA’s \href{https://www.isda-africa.com/virtual-agronomist/}{Virtual Agronomist} building on their knowledge base to provide tailored nutrient plans and agronomic advice based among other sources on their soil fertility data
    \item Samagra’s \href{https://samagragovernance.in/samagrax/}{AMA Krushi AI} that support farmers in Odisha with actionable insights
    \item KissanAI’s \href{https://kissan.ai/chat}{Kissan Copilot}, a voice-enabled, multilingual farmer assistant that provides information about crop management, pest control, weather forecasts and market prices
    \item CIMMYT’s \href{https://www.cimmyt.org/es/noticias/what-is-agrotutor/}{Agrotutor} aiming to provide real-time and localized advice to boost productivity, yield and income of smallholder farmers
\end{itemize}
There are various other initiatives beyond this, including, for example, the recent collaboration of Rwanda's Centre for the Fourth Industrial Revolution with GIZ to bring AI-based advisory to smallholder farmers in Rwanda.

\section{General architecture: How do AIEP advisory systems work?}

\begin{figure*}[tb]
  \centering
  \includegraphics[width=0.9\linewidth]{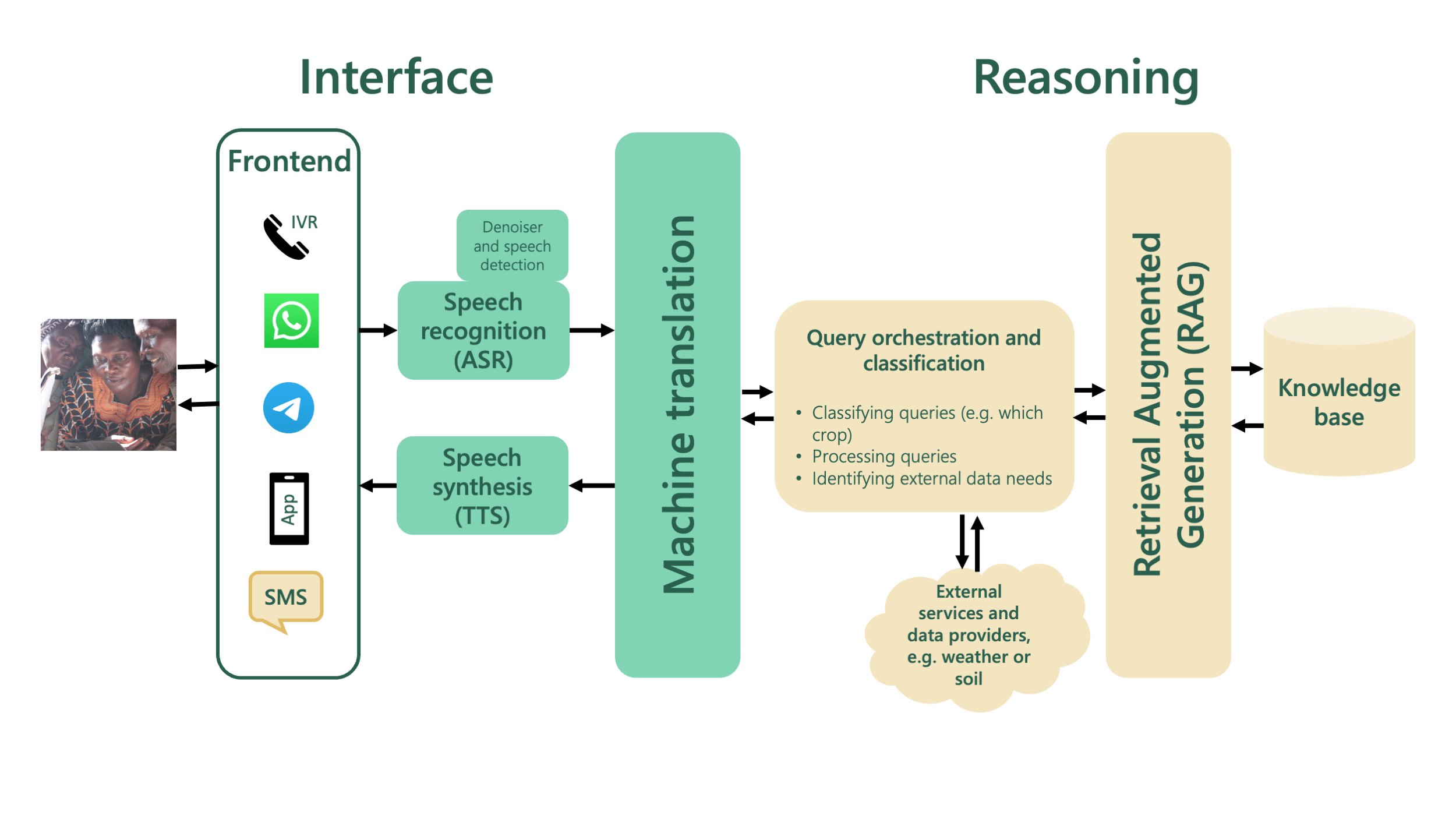}
  \caption{Common high-level architecture of AIEP MVPs}
  \label{fig:aiep_architecture}
\end{figure*}

The cohort developing MVPs under the AIEP initiative built on top of a common high-level architecture (see Figure~\ref{fig:aiep_architecture}), although with differences in specifications, used AI models and a technology stack. This architecture has two major components, consisting in turn of various modules: An interface and a reasoning component. From a software development point of view, the key programming interfaces are between modules and the differentiation between interface and reasoning component aims to support an easier understanding of the overall architecture.

In principle, the architecture is modular; e.g. different channels work with the various implementations of the back-end, although certain modules require certain specifications. For example, an IVR system requires the responses to be of a certain limited length and style to be easily understood via telephone, while apps allow for multi-media content like images and videos. In addition, the cohorts optimized the performance of their MVPs for their specific implementation, and it is unclear whether there are dependencies between, e.g. the machine translation and Retrieval Augmented Generation (RAG) systems. Especially for low-resourced languages with nonstandardized scripts, interoperability of different parts of the language stack is not always given.

\subsection{A walk through}

To illustrate how the architecture works, we can follow an exemplary question through the system.

Imagine a user calling the number of one of the AI-based advisory services in Embu County in Kenya. Upon being connected, the user hears an introductory message from the service explaining basic functionality and potentially providing an example question. The user is a bean farmer waiting for the rain that has not been forthcoming as usual at this time of the year. Unsure of the right timing to plant this season’s beans, she asks “Mimi ni mkulima wa maharagwe katika Kaunti ya Embu. Ni wakati gani mzuri wa kupanda maharagwe yangu?” (meaning in English “I am a bean farmer in Embu County, when is the best time to plant my beans?”). This audio message is transmitted to the system and first transcribed by an automated speech recognition (ASR) model, resulting in her question in text form. This text is then translated into English using a machine translation (MT) model. The question transcribed and translated into English is then processed by the reasoning component. First, the query is classified as related to beans and requiring weather information which the system is requesting from an external weather information provider. The question and the weather forecast are then handed to the LLM, which identifies the relevant information in the knowledge base, in this case planting advice and calendars for beans in Kenya depending on the weather forecast. Based on the forecast and the relevant information identified, the LLM then creates a response to the farmer. This response is then translated into Swahili by the machine translation (MT) model, before the resulting Swahili text is turned into an audio file by a Text-to-Speech (TTS) model. This results in a Swahili audio file that is then sent back to the user who hears it on her phone between 5 and 8 seconds later.

\subsection{The interface}

The interface, or front-end, combines communication channels with speech technology to enable speech input and output and machine translation as the key to multilingual services. The goal of the interface is to allow for access tailored to the needs of the specific user in terms of literacy, hardware, and language.

Farmers interact with MVPs through various communication channels, including telephone lines (based on existing Interactive Voice Response (IVR) infrastructure), SMS messages, smartphone applications, and messaging platforms like WhatsApp or Telegram, although not all MVPs cover all channels. Combining WhatsApp or smartphone applications for users with smartphone access and telephone lines for users with feature phones have proven the most common choice that balances access for users with limited digital literacy, hardware, and Internet access with the broader options for content that digital channels like WhatsApp or applications offer.

Especially telephone lines and IVR require a well-functional speech stack, consisting at least of an automated speech recognition (ASR, also Speech-to-Text or STT model) model to transcribe incoming speech and a text-to-speech (TTS) or speech synthesis model to create speech from text responses. The dynAg cohort also successfully has added a denoiser to improve ASR performance to this stack, and many cohorts have run speech detection models that are usually smaller and faster and can cut out silence so that the ASR model receives only smaller audio files, therefore accelerating processing times and therefore reducing the latency between question and response.

The AIEP MVPs achieve multilingual services in English, Swahili, and Hindi mostly through machine translation (MT), as the reasoning component is mostly English based, albeit Viamo and Sahaj also successfully built a multilingual RAG system to allow them to leverage Swahili agricultural content in Kenya. With improving LLM performance, the MT module could be integrated with the LLM in charge of reasoning\footnote{As reasoning is used loosely in current discussions around LLMs, it might be helpful to define our use: We understand reasoning broadly and encompass both what is traditionally referred to as natural language understanding (NLU) and natural language generation (NLG). Therefore, it refers to all processes between the question as input and the response as output if not otherwise specified, and not to the specific meaning of reasoning in logic as drawing conclusions from facts and premises.}. It therefore sits on the edge between the interface and reasoning components. We list MT for the moment under the interface components as its task is to broaden the access and, therefore, the capabilities of the interface. 

\subsection{The reasoning component}

The reasoning component aims to create correct, relevant, actionable, and easily understood responses to farmer questions. This requires four key tasks:

\begin{enumerate}
    \item Understanding and processing the question
    \item Identifying external data required
    \item Querying the knowledge base for relevant information
    \item Generating a response based on 2) and 3)
\end{enumerate}

Tasks 1 and 4 are traditionally referred to as natural language understanding and generation, respectively, but by now are covered by single Large Language Models (LLMs).

To fulfill these tasks, most cohorts relied on four key modules:

\begin{enumerate}
    \item A query orchestration and classification module
    \item A set of external services providing further data e.g. weather forecasts, soil information or market prices
    \item A LLM-based Retrieval Augmented Generation (RAG) system to generate a response based on information retrieved the knowledge base
    \item The knowledge base
\end{enumerate}

In addition to substantially improving understanding and generation by LLMs, the improvements in usability in recent years stem from integrating LLMs with external services. At the start of the MVP development process, this integration was just emerging with OpenAI’s first publication of LLM function calling\citep{OpenAI2023FunctionCalling}\footnote{Function calling describes fine-tuning models to both identify when a predefined external service is required and to output structured output which another program can use to request information from the external service.}. The AIEP cohort therefore first developed versions of a query orchestration and classification module external to the LLM to improve performance by identifying relevant crops and to classify whether external information like weather or soil data was needed and to then request this weather or soil data. Through this module, AIEP cohorts increasingly added information provided by external services, mainly weather data, soil maps for Kenya provided by \href{https://www.isda-africa.com/isdasoil/}{iSDA} and where available market prices, to the responses.

With further improvements in integrations between LLMs and external services, such as function calling, these modules are likely to become obsolete. More recent developments like the Model Context Protocol (MCP) \citep{AnthropicMCP2024} will further contribute to this integration and allow further acceleration of how AI-based advisory services can integrate external services.\footnote{Model Context Protocol (MCP) was recently introduced by Anthropic. It aims to unify how LLMs interact with external services (function calling only improves the output of LLMs, but does not provide a common interface). One potential challenge of custom integration, e.g. via function calling, is the exponential rise of required integration when N AI solutions each have to define integrations with M external data sources, therefore substantially slowing down progress. At the same time, current solutions allow only one-off requests, while MCP allows continued sessions and interactions between LLM and the external service} To support the application of these developments in agriculture, Digital Green has recently launched an adapted version of MCP for the agricultural sector called AgMCP (Agricultural Model Context Protocol).
The further integration of tasks into LLMs also applies to the language tech component of the MVPs, where future work likely will explore whether those models also could be used to take on the speech and machine translation tasks, therefore reducing latencies and integrating many functions of the interface and reasoning components into a single model (see more on latency in the section below).

One key external information for the LLM are the knowledge bases of agricultural information. Most cohorts relied on the now commonly applied and controllable Retrieval Augmented Generation approach (RAG, see more on evaluation of LLM capabilities in agriculture below). Retrieval-Augmented Generation (RAG) \citep{Lewis2020RAG} combines a large language model with a retrieval system to improve the accuracy and relevance of generated responses. Instead of relying solely on its internal knowledge, the LLM dynamically retrieves relevant documents or facts from the knowledge base. RAG improves transparency and gives implementers more influence over the information an LLM uses, but it is not a full safety mechanism. Retrieval mistakes and hallucinations during generation can still lead to incorrect or unsafe advice \citep{niu-etal-2024-ragtruth}.

Building on RAG frameworks, the knowledge base is a key component of the solution. Most cohorts included dozens to more than 100 different sources of agricultural information, often manual and guidelines provided by research institutes such as CGIAR globally or KALRO in Kenya or government agencies and international organizations. Accessing and validating the content for this knowledge base remains a very manual and time-consuming process. 

\subsection{Learnings}

\subsubsection{Interface}

The experience of the AIEP cohorts highlights the ongoing trade-off between broadening access and providing rich, user-relevant information. On the one hand, using channels that farmers already use, such as WhatsApp, lowered barriers to adoption and fostered regular engagement, particularly among younger or more digitally literate users. However, to reach marginalized groups, such as women farmers or the elderly, who often face constraints in digital literacy, hardware access, or familiarity with new technology, telephone line systems proved indispensable. This confirmed the importance of a multichannel strategy that adapts to varied user needs and local realities.

The experience of the Digital Green cohort offers concrete insight into these trade-offs. They initially launched their service on Telegram because of the ease of development and lower costs than comparable services. Although Telegram is less commonly used than WhatsApp, Digital Green's user strategy is based on extension agents and other multipliers that they train so that the onboarding process is less of a challenge. However, rural users tend to be highly visual; they connect more easily with images, videos, and catalogs that can be accessed with simple taps. The typing or recording of queries often introduces friction, especially for first-time users. A visual first experience reduces this barrier, making it easier for users to engage and build trust. As that trust grows, many may naturally progress to asking queries and using more advanced features. For this reason, the cohort transitioned to a dedicated Android app to offer richer, more intuitive engagement options. A mobile app also simplifies the discovery of value, users can more easily explore what the platform offers through structured navigation, push notifications, and contextual cues, which are harder to implement in messaging platforms.

Language technology also plays a pivotal role in shaping access. While service provision in English, Hindi, and Swahili was relatively straightforward, extending these solutions to lower-resourced languages such as Gikuyu in Kenya or Bhojpuri in Bihar continues to present considerable obstacles. For many of the most vulnerable users, including nonliterate farmers and those speaking minority languages, language access is the cornerstone of inclusion, but supporting these languages is difficult both technically and operationally. Even if parts of the language technology stack like ASR or MT exist, integrating them for languages with nonstandard scripts or limited digital resources remains a hard challenge, often resulting in insufficient performance to provide services in these languages. 
Even for languages like Hindi, which are generally well covered, some edge cases which are frequent in agriculture, but not general use of language technology continue to cause errors. For example, ‘mushroom’ 
may be pronounced as ‘mushraum’ 
in Hindi which was transliterated as 
'machchharon' (which is phonetically similar) which in turn got translated to mosquitoes in the English translated prompt, triggering a wrong query for the AI search engine resulting in a non-relevant response. For machine translation, it causes problems that paddy can be spelled as dhaan or dhan. Google and Bhashini MT models, among the best existing ones, interpret some of these variations as 'dhuhn', which means 'money' instead of 'paddy'.

Current system performance is robust for high-resource languages, but the lack of scalable, high-quality solutions for others continues to exclude some user groups. Addressing these challenges will require not only technological advances but also targeted investment and partnerships for language data creation and language models. 

\subsubsection{Latency}

Latency is the time between a user sending a question and receiving a response. This has been a consistent challenge for all AIEP MVPs, as well as related solutions that all struggle to consistently reduce latency to below 5 seconds for speech-based services.\footnote{We invite the reader to slowly count to 5 to get an impression of what this implies for the user experience.} One of the key open questions remains which part of the architecture drives this latency and, therefore, which optimization would lead to lower and more conversational latency. Those drivers do not purely stem from the interface component, but the language AI deployed there likely has a substantial impact, and latency is crucial as it affects user experience.

The number of model inference calls – at least four consisting of speech recognition, machine translation, LLM, and speech synthesis – might be one driver, and therefore moving more tasks to the LLM is a current topic of testing. For example, state-of-the-art LLMs can successfully combine translation and reasoning for some languages like Swahili and Hindi. In addition, recent iterations of LLMs integrate voice support directly into the model and API, for example, GPT-4o or Gemini's real-time audio APIs.

If not driven by model inference (or other software-inherent processing times), network latency is the second key driver of overall latency. Some of the AIEP cohorts have indeed achieved substantial improvements in latency by moving toward model providers or model hosting solutions located in the country (given the infrastructure available primarily in India) and reducing the size of the audio files sent from the user’s device to the cloud-hosted application. One cohort, for example, managed to reduce average latency by a factor of 25 to just below 10 seconds through these measures.

Hosting models on user devices, e.g. small-scale speech models or even whole reasoning models, would be the most thorough solution for network latency. However, the trade-off in terms of performance but also a potential longer inference time of those smaller models remains unclear. Such a solution would at the same time solve users' potential connectivity challenges and also reduce the costs from sending large audio files, which need to be paid for either by the user or the solution provider.

\subsubsection{Reasoning Component}

In the back-end, the development of comprehensive and contextually relevant knowledge bases demanded a significant commitment of time and expertise from all cohorts. Most high-quality and up-to-date agricultural information is available only in English, requiring robust machine translation pipelines or LLMs with advanced multilingual capabilities to support non-English-speaking users.

However, translation alone cannot compensate for content that is insufficiently localized, missing key crops, or lacking practical advice tailored to regional farming contexts. Identifying and accessing new content and curation, validation, and ongoing maintenance of content proved to be continuous processes throughout the development phase, not one-off tasks. Usually, cohorts needed to identify and engage external partners, which caused substantial work given that an overview of what agricultural information is available is not centrally available, and good networks and partnerships are required to identify and get access to partners with this information. In case of proprietary information, the process often required setting up agreements to access the information. Even in the case of openly available data, cohorts needed to look through content with qualified staff to ensure its quality and relevance to avoid serving outdated, wrong, and irrelevant information. Given the fragmented landscape of agricultural data and content, cohorts also duplicated much work, underscoring the importance of partnerships and better data sharing.

Although LLMs brought flexibility and the ability to respond in a variety of formats, they also introduced new risks, especially around accuracy and hallucination, while our evaluation capabilities are only emerging. Here, the use of Retrieval-Augmented Generation (RAG) frameworks emerged as a best practice, allowing partners to more finely control the content underlying an LLM's response. This remains particularly important, as wrong advice can cause substantial material harm to farmers and as governments rightly closely scrutinize the quality of AI-based advice. As AI-based advisory increases to rely on knowledge learned by LLMs during pretraining or fine-tuning versus querying external knowledge bases via RAG, it remains important that evaluation ensures that the advice is factually correct and does not risk causing harm.

The options for integrating external services are evolving rapidly with function calling being refined, or the Model Context Protocol continuing to see rising adoption. At the same time, RAG continues to be the standard for managing the content which AI-based advisory is providing to smallholder farmers, and how to integrate those two trends will likely be a key question in the future.  

\section{How to approach building AI-based advisory?}

Given its fortunate timing to have started just with the release of ChatGPT, the AIEP initiative was among the first to develop AI-based agricultural advisory services and therefore generated many learnings on how to approach developing such systems in agriculture and other domains. In this section, we provide an ordered list of questions that partners should investigate to identify obstacles early and reduce development time.

\subsection{The Team}

The AIEP MVPs were built by multidisciplinary teams across various organizations because building tailored and impactful advisory solutions requires the combination of different skill sets that only very few organizations combine successfully. Technically, besides general software development skills, in-depth AI knowledge is required to choose the right models, frameworks, and successfully develop, test, and monitor the AI modules of the application. When working in low-resource languages, it is also necessary to work directly with native speakers of those languages, and having those speakers on the team massively accelerates the development process.In-depth agricultural knowledge (as in every domain) is necessary to succeed. Lastly, an experienced product team should be charged with coordinating those different expert skills and ensuring that the overall application meets the users' needs in a financially sustainable manner. 

\subsection{Monitoring and measuring}

\begin{figure*}[tb]
  \centering
  \includegraphics[width=0.9\linewidth]{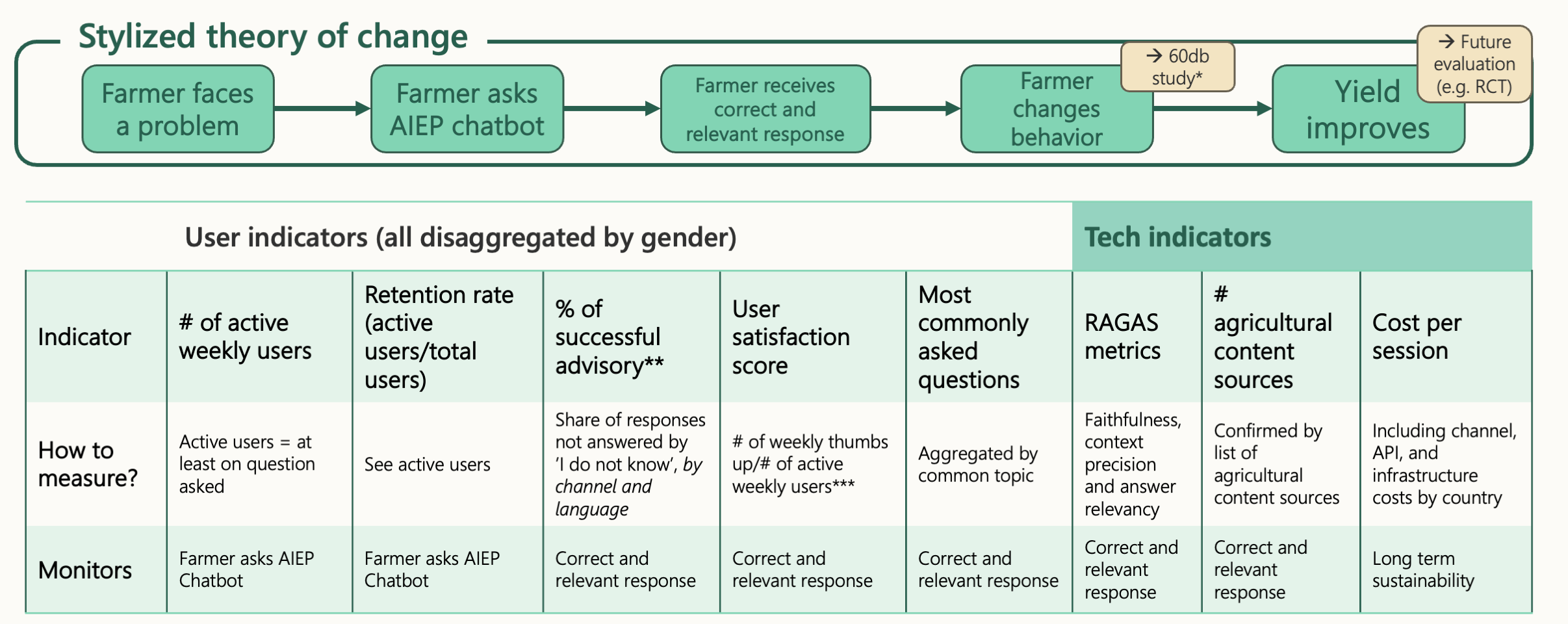}
  \caption{Stylized theory of change and AIEP Standard Indicators}
  \label{fig:aiep_indicators}
\end{figure*}

Like all development projects, building and maintaining AI-based advisory needs and gains from good monitoring and consistent measurement to have data sources that allow to identify sources of errors, understand user acquisition and prepare impact measurements.
As an initiative consisting of multiple teams, the AIEP initiative created a set of standard indicators (see Figure~\ref{fig:aiep_indicators}). Those indicators were collected, except for some data points that were not feasible, by all cohorts, allowing for comparison between the different teams. 

\subsection{Content: What information do you want to provide?}

While AI-based agricultural advisory systems have the potential to provide much more timely, personalized, and relevant information to smallholder farmers, the information they provide will always depend on the content sources which we provide to the AI systems. Even with RAG systems, these content sources still need to be vetted for correctness and relevance, which has proven a substantial obstacle and source of effort.
Therefore, the first step when building such advisory systems should be to verify and plan access to content sources. In agriculture, these content sources can include

\begin{itemize}
    \item agronomic material as provided by research institutions like the CGIAR network globally (see the \href{https://www.ifpri.org/project/generative-ai-for-agriculture-gaia/}{GAIA project}) and organizations like KALRO nationally (e.g. through their \href{https://kadp.kalro.org/home}{Kenya Agriculture Data-Sharing Platform}) or by ministries
    \item databases like crop maps, soil maps (like the one provided by \href{https://www.isda-africa.com/isdasoil/}{iSDA}) or databases of seeds and livestock breeds or government schemes
    \item market price information including on inputs like seeds and fertilizers and on to-be-sold crops
    \item weather information (e.g. provided by \href{https://tomorrownow.org}{TomorrowNow} or national meteorological institutes e.g. in India or Kenya) and other geospatial data
\end{itemize}

As accessing this data is a key factor in providing high-quality advisory services while also often depending on partnerships and potential contractual arrangements, this should be the starting point when building AI-based advisory systems. For quick deployment, an evaluation of inherent capacities of LLMs might provide a substitute, but we expect that successful advisory solutions in agriculture as in other domains will eventually need access to high-quality and relevant content, so should plan for this early.

\subsection{Channels: How will your users interact with your advisory service?}

Once the content is clarified, the user interface or channels should be considered, as their choice will partially determine the technology stack.
In general, the AIEP experience shows that a multi-channel approach, e.g. combining WhatsApp with a telephone line, is necessary to balance access with information richness, despite the efforts required to build and maintain multiple channels. These insights arose from the fact that smallholder farmers are diverse in terms of access to hardware like phones, their literacy and fluency in using the hardware, the languages they speak, and their preferences. Furthermore, cost is a significant factor, as some APIs like WhatsApp Business have stark regional differences. Telephone lines also potentially need local infrastructure or partnerships that need to be scoped. Finally, local regulations also determine how different channels might be used with governments at times applying different regulations to different channels (for example, in India bulk SMS comes with substantially stricter regulations vs. bulk WhatsApp messages).

However, as e.g. Digital Green’s and Tech for Her’s experiences show depending on the users and kind of outreach, this can be adapted. If the primary channel for information includes multipliers like extension agents, or e.g. in health, community health workers, who can more reliably be equipped with smartphones, it might be possible to focus on a single channel tailored to this specific group.

\subsection{Tech: Build on what is already there!}

Developing AI-based agricultural advisory systems can be expedited using existing technologies and frameworks. Open-source platforms, pre-trained models, and established APIs offer a foundation that can be customized to specific needs. Inspired by the basic thought behind Digital Public Infrastructure and Digital Public Goods, AIEP and many related initiatives are currently creating shared resources and experiences that related initiatives can leverage. The goal should always be not to duplicate work and to deploy solutions to users as fast as possible. Building upon these existing technologies, development time and resources can be significantly reduced, allowing teams to focus on tailoring solutions to local contexts and user needs.

On the tech side, while there is not yet a detailed unified architecture beyond the high-level one presented, and might not arise soon given the speed of technological change, various pieces are available open source from models like ASR and MT models to RAG frameworks. The most successful example of building on existing tech is the example of the fifth AIEP cohort consisting of Opportunity International, Safaricom Digifarm and Gooey.ai. Opportunity International had been working on developing AI-based information systems, but was able to substantially accelerate their development first in Malawi and then in Kenya in cooperation with SafariCom Digifarm by partnering with Gooey.ai who were able to reuse what they had in part already developed in cooperation with Digital Green as part of the AIEP initiative.

On the experience side, we hope that papers like this and exchange formats that will follow ensure that mistakes and dead ends can be avoided.

\subsection{User feedback: Deploy early}

Early deployment of AI-based advisory services is crucial for iterative development and user-centric design. Launching a minimum viable product (MVP) allows real-world testing, user feedback, and identification of practical challenges that may not surface during development. This approach enables teams to refine the functionalities of the system, improve user interfaces, and adapt the content to better serve the target audience. Moreover, early deployment fosters stakeholder engagement, builds trust with users, and provides valuable insight into user behavior and preferences, all of which are essential for successful scaling of the service.

\subsection{Think about costs at scale}

Understanding and planning for costs associated with scaling AI-based advisory services is essential for sustainability. Although initial development may be manageable, scaling introduces additional expenses, such as increased server usage, data storage, API calls, and maintenance. For example, integrating services such as weather data, market prices, or language translation can incur variable costs based on usage. Crucially, user acquisition costs need to be sustainable even at higher user numbers and after initial funding has run out.

It is important to analyze these costs early on, explore cost-effective alternatives, and consider partnerships or open-source solutions to mitigate expenses. Furthermore, evaluating the cost-benefit ratio of different channels (e.g. SMS, IVR, WhatsApp) can inform decisions that balance reach, effectiveness, and affordability.

\section{Enablers: Context factors that make building this easier}

The AIEP initiative was always designed as a collaborative undertaking, combining various cohorts trying to solve the same problem to collaboratively explore more options and gain from exchange and sharing. As part of this and related work by partners, we also identified various enablers that can impact the whole ecosystem, making it easier for each individual actor to provide high-quality services. These therefore should be provided as Digital Public Goods (DPGs) or Digital Public Infrastructure (DPI). The key enablers which we identified are:
\begin{itemize}
    \item Data sharing infrastructure
    \item A common corpus
    \item Benchmarking and evaluation
    \item Better language technology
\end{itemize}

We describe these enablers in the context of smallholder farming, but they either apply analogously to other sectors (like data sharing or evaluation) or are directly applicable across sectors (like language technology).

\subsection{Data Sharing}

Access to data that includes not just a corpus of agricultural information, but also relevant language data to fine-tune models or conversation logs to better understand farmer needs is a key enabler to build better AI-based advisory services.

Yet, our experience in AIEP was that data sharing is often hindered by a lack of standards which make communicating about what data is available in which format and quality, a lack of resources to curate data to be shared (to ensure they are as useful as possible for others) and a lack of incentives as some partners might worry about their position.

To alleviate the lack of resources and support easier sharing through a common interface, the AIEP initiative piloted its own data platform supported by Digital Green and built on top of their FarmStack protocol. A common data platform allows for easier sharing and crucially for easier identification of other datasets which could be commonly described. AIEP therefore piloted this activity to address the concrete challenge of the AIEP initiative, but also the sector-wide challenges of fragmented agricultural data and inefficient resource utilization. We aimed to test out in the smaller and coordinated space of the AIEP initiative what might be able to contribute to wider data-sharing activities in the future. By deploying an instance of FarmStack, an open source data sharing interface by Digital Green, the data sharing interface aims to create a unified platform for pooling and sharing tagged, annotated, and curated datasets, modular code repositories, and fine-tuned agricultural AI models. By having a unified platform, we hope to reduce the effort needed to share the various resources needed to build an AI based advisory. 
The incentive problem is more difficult to overcome and likely will require a combination of trust, partnerships, and a common agreement on which resources are common goods and which might be private and potentially part of competing offerings.

\subsection{Common Corpus}

Our experience clearly shows that access to high-quality agricultural data is the second key enabler for high quality, relevant, and personalized AI-based advisory. Although AI systems incorporate increasing amounts of agricultural knowledge (see more on evaluation below), a targeted and easy-to-control advisory service probably still relies on RAG-based systems, although this might change with the advent capable agentic systems. These systems can create high-quality conversational experiences and ease access to agricultural information, thus leveraging investments into this corpus.

As all AI advisory solutions rely on this corpus, this is a good candidate for a Digital Public Good (DPG), as it allows various actors to build advisory services targeted to specific geographies and groups while easily relying on the most high-quality information in a common corpus.

Currently, accessing and curating this agricultural information requires substantial amounts of time and effort, slowing down development and innovation, and leading to repetition of something that is a common challenge. Projects like the Gates Foundation supported GAIA project implemented by CGIAR, CABI, and Digital Green work on solving this common challenge.

\subsection{Better language AI}

As a concrete application of the recent advancements in AI, AI-based advisory will also continue to benefit from further progress in technology. But there are two key areas which likely will substantially improve AI-based advisory while at the same time might not receive sufficient private-sector investments due to their low commercial value.

First, LLMs that have better agriculture and specifically smallholder agriculture in LMIC related capabilities will improve the advice given. These better capabilities can include various pieces, among others:

\begin{itemize}
    \item Better vocabulary that is robust for the different and often very local ways in which crops, diseases and pests and actions are described by farmers 
    \item Better capacity to track available information and gaining the ability to ask questions back to the user or request external information if some is missing
    \item More contextualized agricultural knowledge 
\end{itemize}

Most LLMs are currently used in English and other major languages. Similarly, the AIEP MVPs also mostly run an English LLM-Reasoning Component and create multilinguality through dedicated machine translation models (although those partially are LLMs themselves). This drives up latency and increases costs for development and maintenance, but despite the recent advances of LLMs for low-resourced languages, they still perform poorly for very low-resourced languages which are of crucial importance to provide access to marginalized groups (to get a sense of language use, see \href{https://clearglobal.org/language-maps-and-data/}{CLEAR Global’s language use platform}).

\begin{figure}[tb]
  \centering
  \includegraphics[width=1\linewidth]{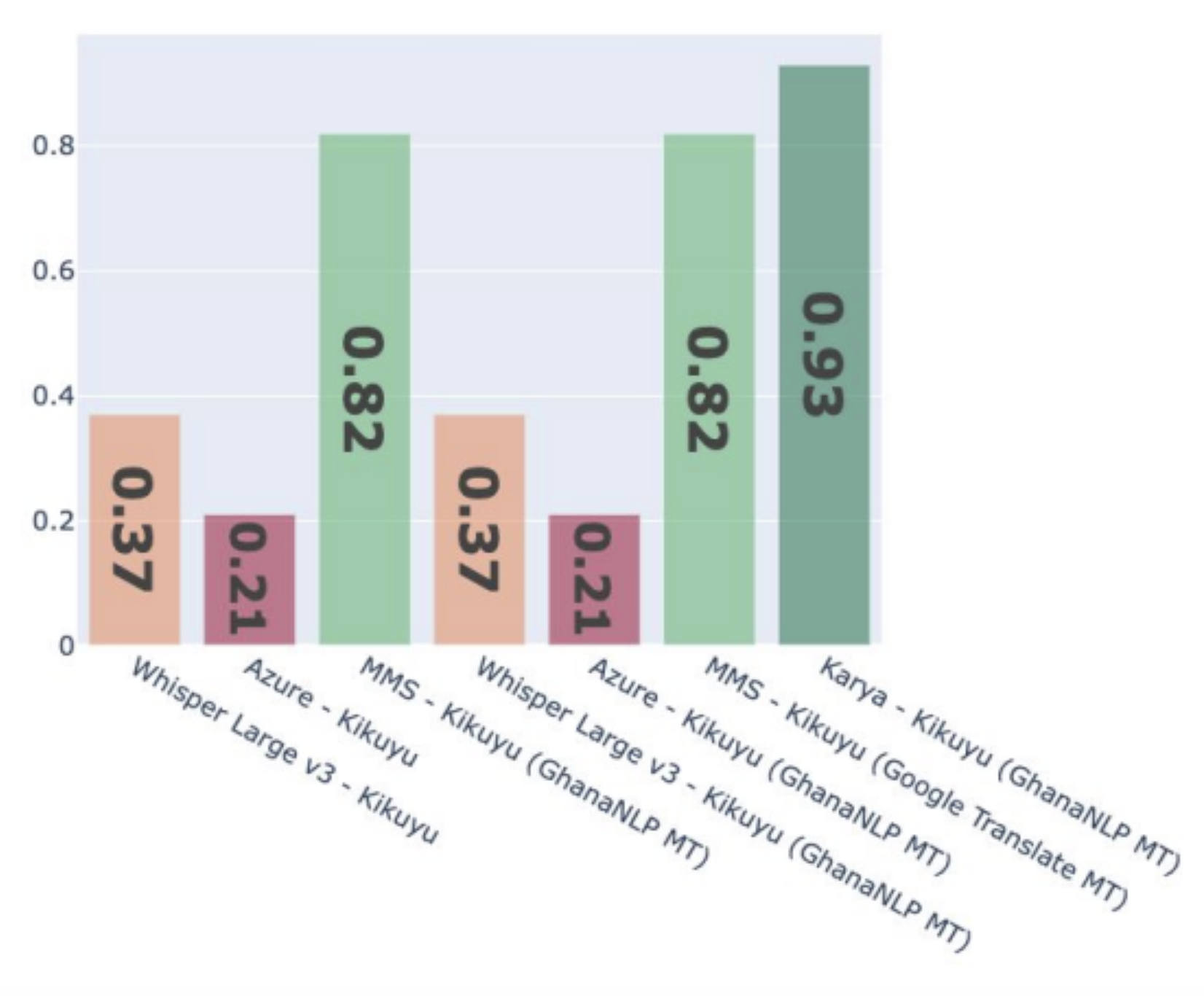}
  \caption{ASR Performance in Gikuyu as evaluated by Gooey.ai}
  \label{fig:gikuyuASRperformance}
\end{figure}

Further improvement for low-resourced languages for all parts of the language technology stack, i.e. speech recognition, translation, speech synthesis, and reasoning, will therefore improve the access and quality for an ever broader but also often marginalized group of farmers.
The experience of Digital Green, Karya, and Gooey.ai in creating ASR for Gikuyu shows that this is achievable, but also requires a collective and concerted effort. The teams collected 275 hours of voice samples from more than 500 farmers of which 76 hours were validated (with a pass rate of 81\%). The ASR model (a fine-tuned MMS model) trained on this data showed very good performance, constituting the best available Gikuyu ASR model at that time and illustrating what is possible (see Figure~\ref{fig:gikuyuASRperformance}). However, as the architecture shows, we need a functional combination of ASR and machine translation to use this technology in applications.

\begin{figure}[tb]
  \centering
  \includegraphics[width=1\linewidth]{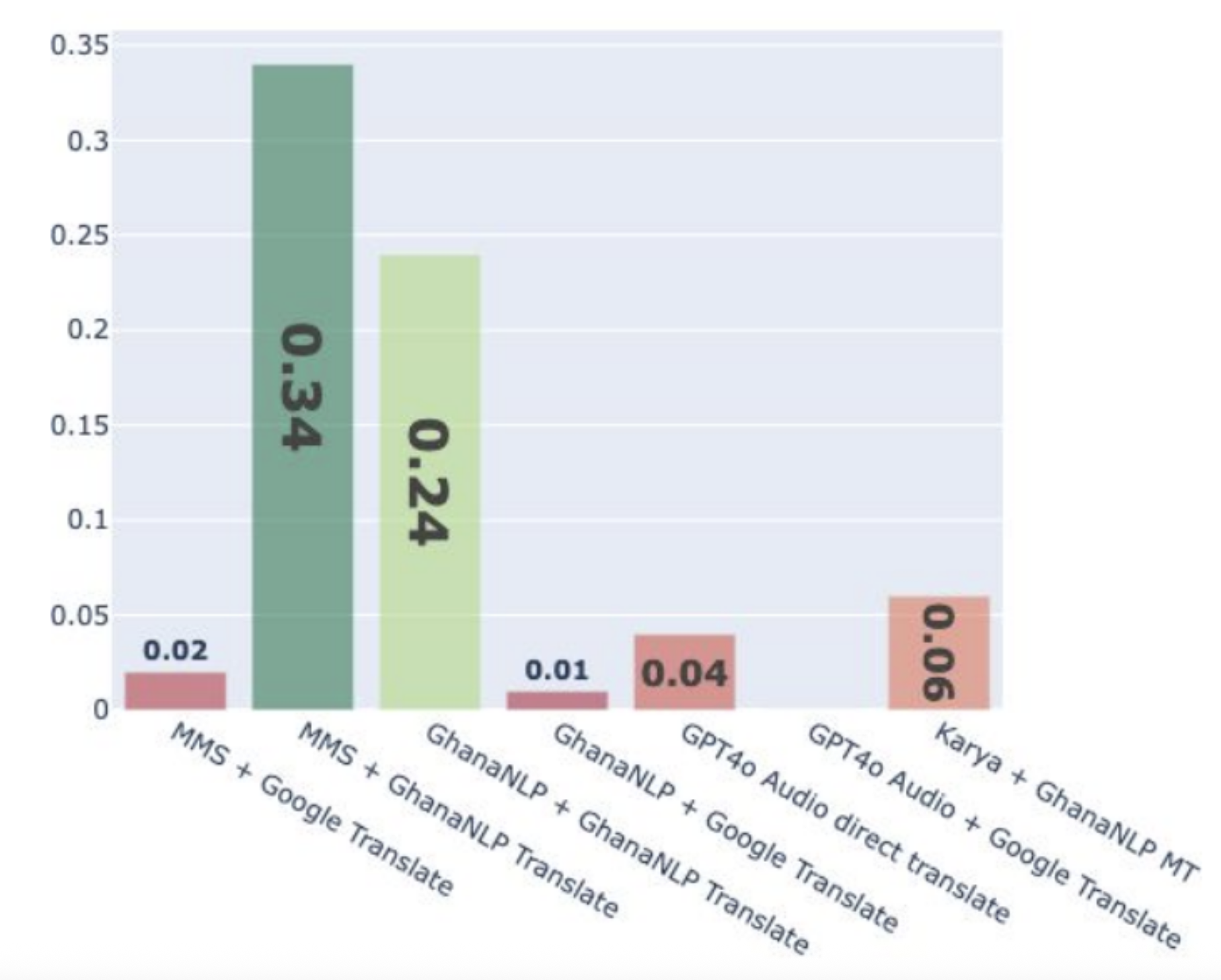}
  \caption{Combined Performance of ASR and MT in Gikuyu as evaluated by Gooey.ai}
  \label{fig:gikuyuASRandMTperformance}
\end{figure}

Performance is substantially poorer (see Figure~\ref{fig:gikuyuASRandMTperformance}), which shows performance evaluated through an LLM with 0 meaning "nonsense / no meaning preserved", 1 meaning "perfect meaning and grammar" and 0.4 meaning "some meaning preserved". This might be due to how models handle diacritics in Gikuyu (diacritics are signs written above or below letters indicating a difference in pronunciation). Gikuyu uses  a “$\sim$” diacritic on u's and i's. However, this is at times transcribed with a “–“. The golden evaluation dataset only used “–“ while some models transcribe either exclusively into “$\sim$” or add additional “–“. Native Gikuyu speakers can ascertain the difference. However, the models themselves are sensitive to these differences and generate substantially different translations depending on the presence or absence of different symbols for diacritics. In addition, Gikuyu, like many low-resource languages, does not have a standardized writing system to which we could refer (see also \citet{Chen2024InterplayMTDiac}). 

\subsection{Evaluation and benchmarking}

Although LLMs and other language AI is heavily evaluated, we have only limited insights into the performance of this technology in low-resourced languages and for social impact use cases. \href{https://lmarena.ai}{ChatbotArena}, one of the standards for evaluating LLMs, does not provide detailed information for LLM performance in low-resourced languages or domains of interest in the development space such as agriculture or health. While we have various evaluation sets and data on LLM performance in English and in domains like programming, math, or law, teams developing AI-based advisory struggle to identify the best working models and to judge whether new models have the potential to improve the advisory services, thereby justifying the effort of integrating them into an application.

At the same time, we need different measurements for different stages of technology implementation. While benchmarks and evaluation datasets serve us to assess progress in the foundational technology, we need additional human evaluation (see case study below), user studies (like the 60 decibel survey) and eventually rigorous evaluations of the impact of technology (like randomized control trials which Digital Green has conducted on their previous video-based digital advisory services, see \citet{Baul2024JDE}).

For low-resourced languages, we face not just a lack of training data, but also high-quality evaluation data. Many existing evaluation sets contain errors (see, e.g., \citet{abdulmumin2024correctingfloresevaluationdataset}) or are culturally not representative (see, e.g., \citet{Singh2024GlobalMMLU}), while nonstandardized scripts make the evaluation harder\footnote{Most evaluation approaches assume that there is one correct response which is not the case if there are multiple commonly accepted ways of writing a word.}. Similarly, non-standardized scripts in low-resourced languages make developing the technology more difficult as different parts of a language technology stack might tend to use different styles and integration becomes a challenge (see the Gikuyu ASR case study above).

Better benchmarking and evaluation therefore would be a common enabler, reducing uncertainty about the best available foundational technology, and therefore providing guidance to implementers, but also incentives to foundational technology developers to improve their models to better serve smallholder farmers.

Having identified this common need, the AIEP initiative also started work on evaluation data for AI advisory, called the golden question-answer set or Golden Q\&A to allow for the agriculture-specific evaluation of LLMs.

One dataset was created by Digital Green (see also \citet{DigitalGreen2025AnnualReport}). Based on more than 2.2 million farmer queries from not just Kenya and Bihar but also other geographies in India, Africa, and Latin America, Digital Green curated a dataset where each answer is vetted by multidisciplinary reviewers, guided by annotation protocols that include gender and equity considerations. As of April 2025, ~5,000 curated Q\&A pairs exist for Bihar in English and Hindi, with expansion underway in Odisha, Telangana, Kenya, Nigeria, Ethiopia and Brazil. Multimedia content (voice, images) is also being evaluated now, using Digital Green's repository of 100,000+ voice notes and 25,000 images. Digital Green also defined tiers of dataset maturity:
\begin{itemize}
    \item 10–20K Q\&A pairs yield baseline model quality for core crops/languages.
    \item 50–100K Q\&A pairs enable broader regenerative themes and multilingual resilience.
    \item 200K+ Q\&A pairs unlock specialized, real-time learning for diverse farmer profiles.
\end{itemize}

In addition to Digital Green’s dataset, the AIEP initiative also created a targeted Golden Q\&A to investigate various evaluation approaches. This data set was created in cooperation with various agronomists in Kenya and India, resulting in a data set of 114 quality-assured English questions and answers.\footnote{Further investigation will require expansion of this work into other languages as Digital Green is already supporting. The analysis of additional languages was unfortunately beyond the scope of this work.} It involves questions selected by agronomists and common questions extracted from 28,896 user questions across the AIEP MVPs, as well as reference responses written by the agronomists. In cooperation with agronomists, we created a set of multiple-choice questions which, while less representative, are substantially easier to evaluate. Due to concerns about data leakage, the data set is not publicly released.

To analyze these results, we posed the same questions to a set of various LLMs. We then calculated various automated metrics (e.g., n-gram overlaps and embedding similarities) and employed an LLM-as-a-judge approach where another LLM is asked to evaluate the LLM responses. To assess the quality of those automated metrics, we also asked the agronomists to evaluate the LLM responses across five metrics: Factual correctness, comprehensiveness, relevance, actionability, and intelligibility. We then correlated these human expert evaluations (which we treated as the ground truth) with the automated metrics.

The results showed some variation in performance but also illustrated that creating non-trivial questions with a single correct answer independent of context is difficult. Many answers in agriculture are very context-dependent: What was grown on the plot before, how much fertilizer has already been applied, or which inputs are available and affordable for a given farmer, potentially using a host of different government support schemes, are questions that determine the correct and appropriate answer. At the same time, this context creates substantial complexity that is difficult to represent and requires more detailed data than is currently available.

\begin{figure*}[tb]
  \centering
  \includegraphics[width=0.9\linewidth]{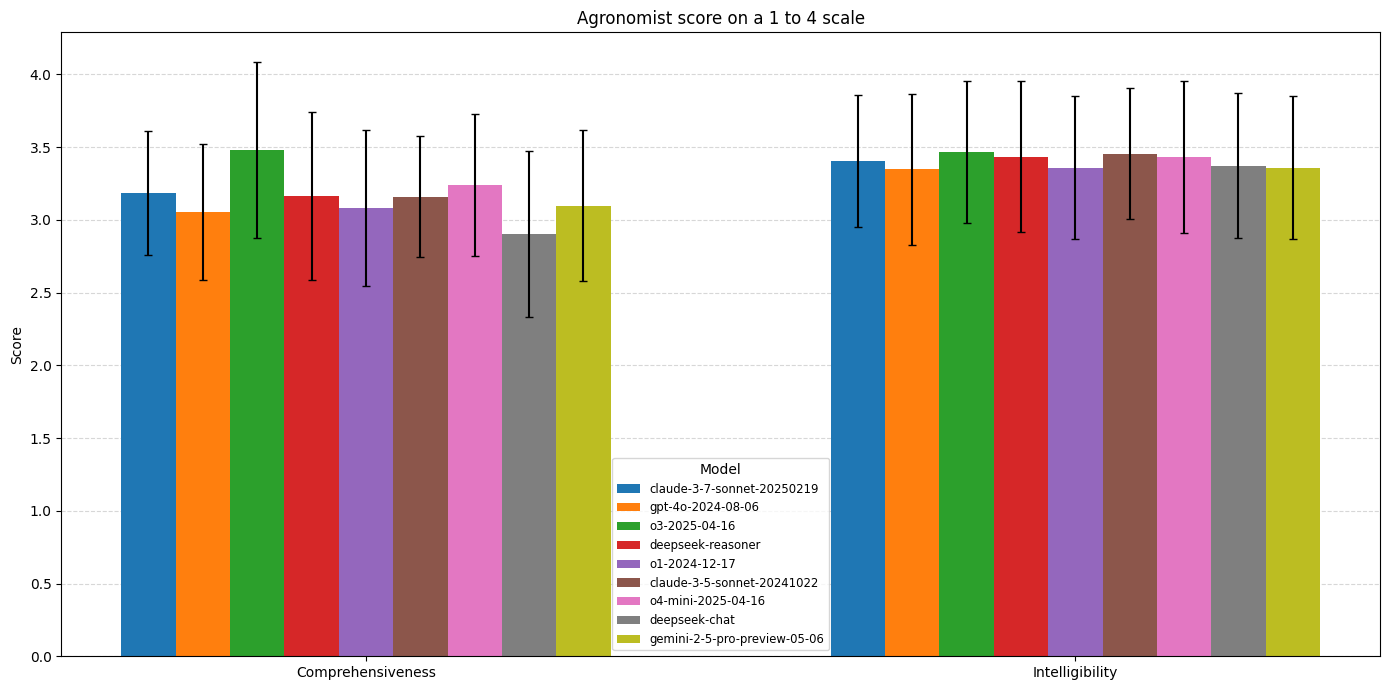}
  \includegraphics[width=0.9\linewidth]{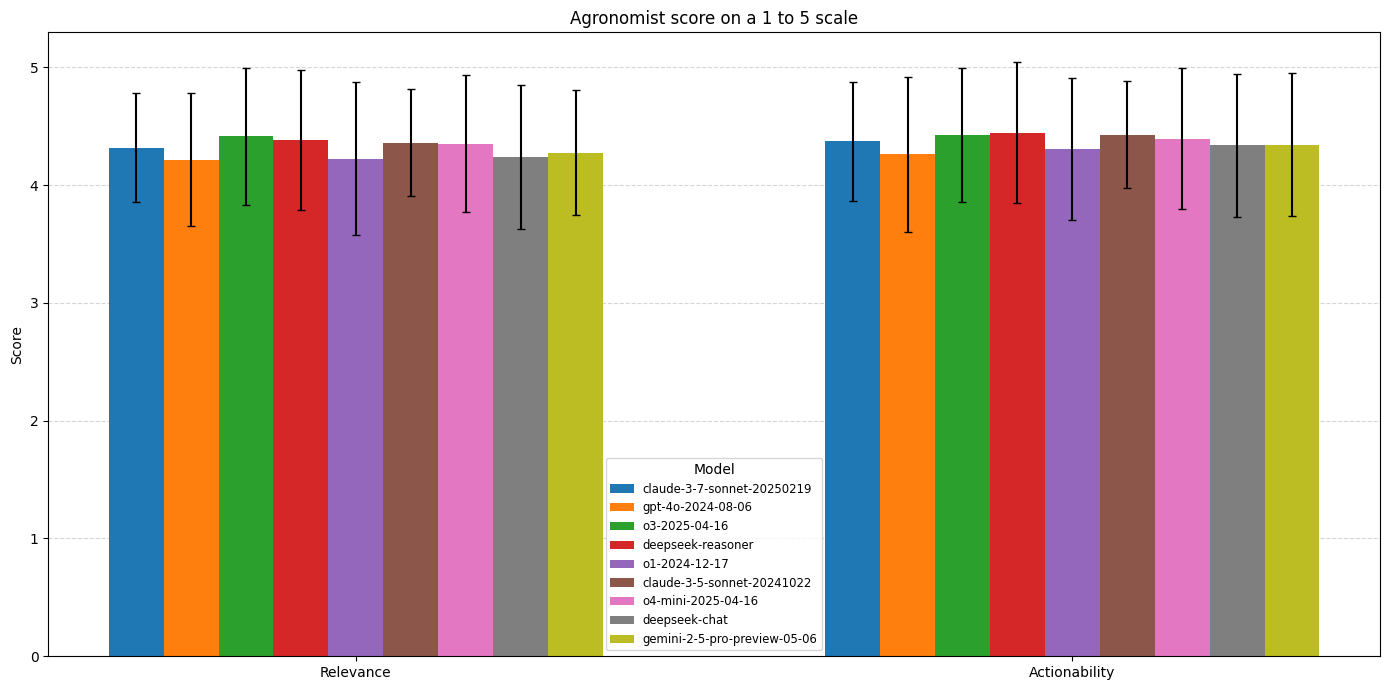}
  \caption{LLM performance on agricultural questions as rated by human agronomists}
  \label{fig:llm_agronomist_eval}
\end{figure*}

The analysis of the open-ended questions proved insightful: First, the agronomists rated the performance of various LLMs very highly, without statistically significant differences between them (see Figure~\ref{fig:llm_agronomist_eval}).

\begin{figure*}[tb]
  \centering
  \includegraphics[width=0.9\linewidth]{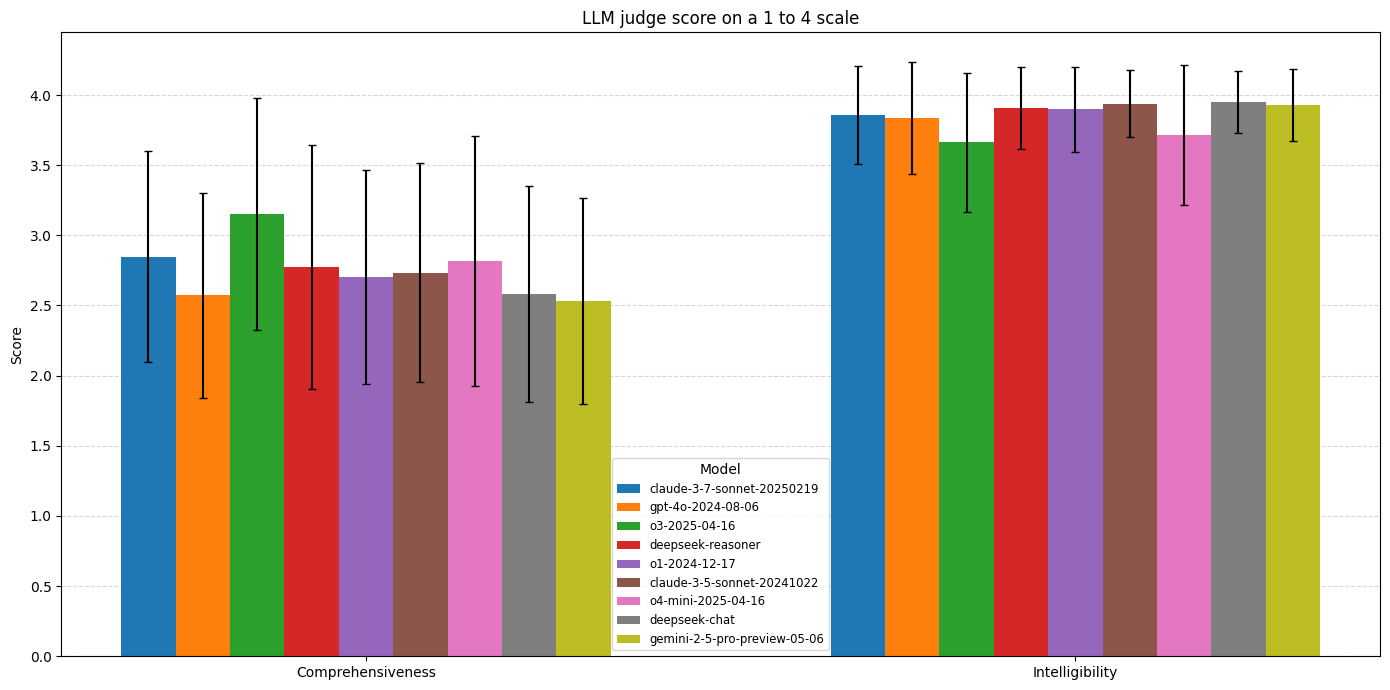}
  \includegraphics[width=0.9\linewidth]{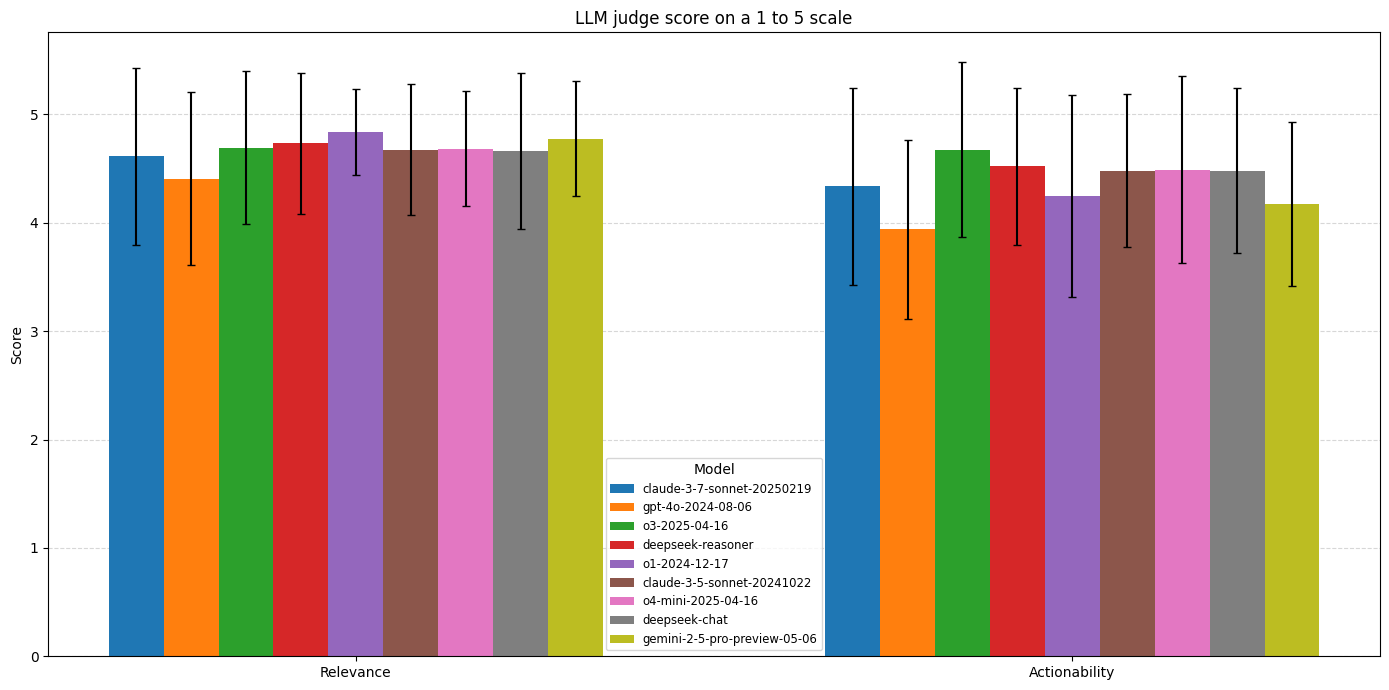}
  \caption{LLM performance on agricultural questions as rated by an LLM-as-a-judge approach}
  \label{fig:llm_judge_eval}
\end{figure*}

We also found that finding good evaluation metrics is difficult. As Figure~\ref{fig:llm_judge_eval} shows, an LLM-as-a-judge approach generally yields similar evaluations to human agronomists, albeit slightly lower at times.

\begin{figure*}[tb]
  \centering
  \includegraphics[width=0.9\linewidth]{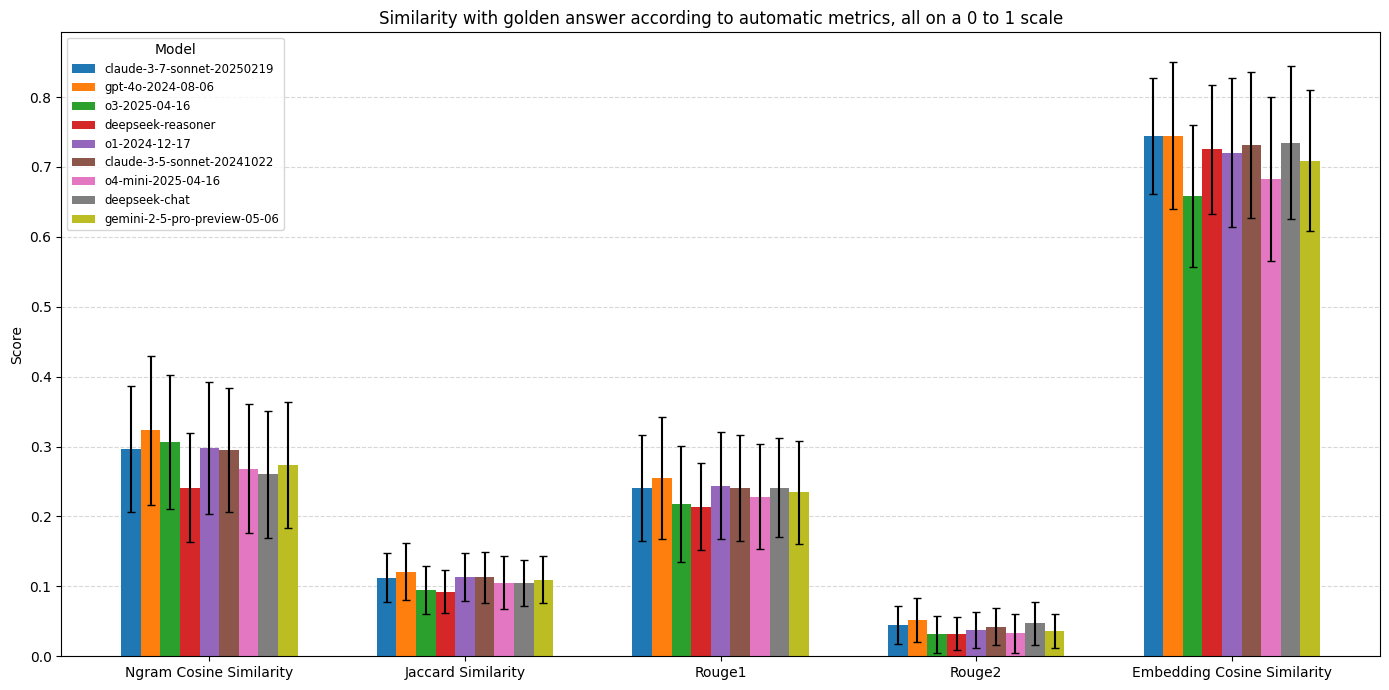}
  \caption{LLM performance on agricultural questions as rated by different automated metrics}
  \label{fig:automated_metrics}
\end{figure*}

Other automated metrics such as n-gram cosine similarity, Jaccard similarity, ROUGE-1, ROUGE-2 and embedding similarity show a much more mixed picture (see Figure~\ref{fig:automated_metrics}).\footnote{N-gram cosine similarity is calculated by breaking strings into overlapping sequences of words (n-grams) and counting the occurrence of each n-gram in both strings. For the overall measure, the similarity of the vectors of the n-gram counts is measured by the cosine similarity. Jaccard similarity also compares n-gram counts, but measures similarity by dividing the number of common elements by the number of unique elements. ROUGE-1 and ROUGE-2 similarly compare overlap in unigrams (1-grams of words) and bigrams (2-grams, sets of two subsequent words), respectively.} For all metrics that compare n-grams, we expected low values given the large number of variants in which a response can be formulated. The embedding similarity provides a more semantically informed picture and greater overlaps.

\begin{figure*}[tb]
  \centering
  \includegraphics[width=0.9\linewidth]{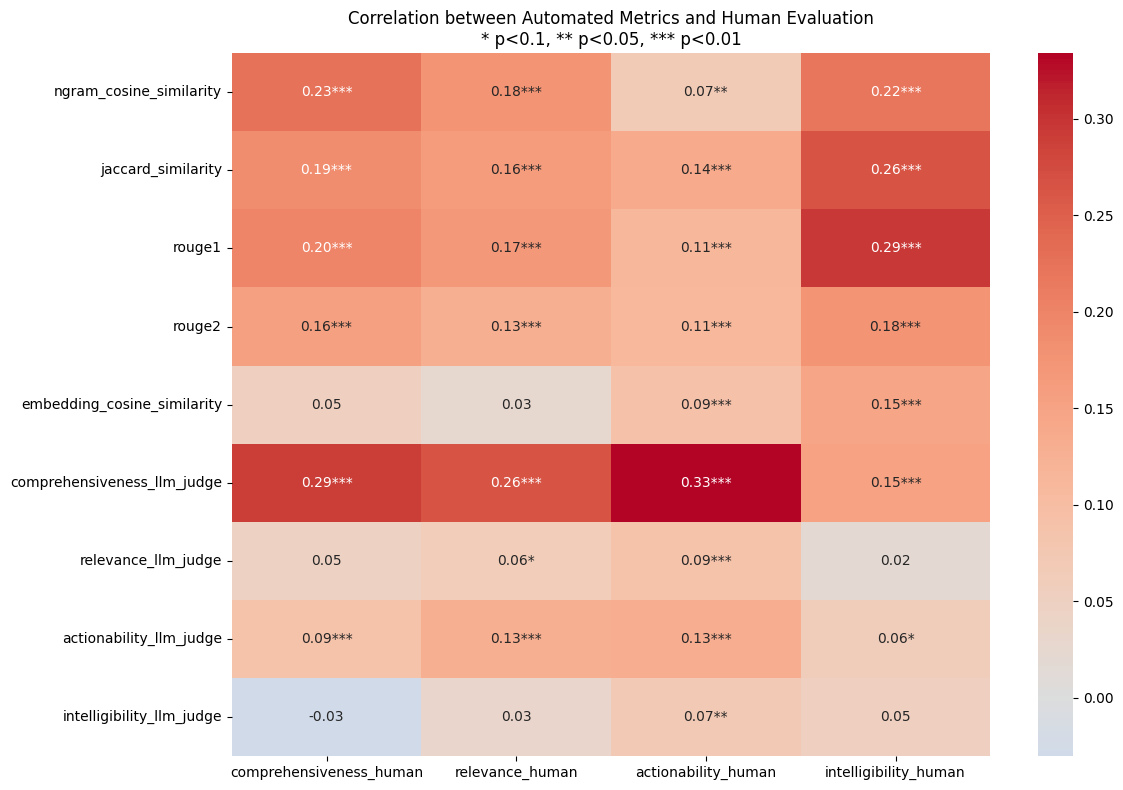}
  \caption{Correlation between automated metrics/LLM-as-a-judge and human evaluation of LLM agricultural performance}
  \label{fig:eval_heatmap}
\end{figure*}

To further investigate those metrics, we correlated the various metrics with the evaluations by human experts. This correlation showed that while various metrics correlate significantly with human evaluations, only an LLM-as-a-judge approach showed at least low to medium correlations on more than one metric (see Figure~\ref{fig:eval_heatmap}). In conclusion, this implies that while especially LLM-as-a-judge can serve as a good substitute for human-expert-evaluation, and could even substitute golden Q\&As, if a broad assessment is required, finer differences between models will require more precise evaluation datasets and metrics. 

\subsection{Learnings on DPGs and DPIs}

Through our work of creating DPGs aimed at aiding the development of similar solutions in the future and our engagement with partners developing the enablers described above as DPGs, we collected insights and learnings about DPGs and DPIs in general. The general premise of DPGs and DPIs remains: Widespread impactful digital technology requires shared investment in public goods and infrastructure that allow us to build this technology quickly, safely, and at scale.

At the same time, the sector is learning about the best ways to design, develop, and maintain those goods and infrastructure and about the pitfalls that lie ahead.

Building robust and scalable DPGs requires substantial initial investment beyond what is required to develop a technology for the use of individual actors, and sustaining them requires ongoing funding for maintenance, upgrades, and community support. For example, the development and continuous improvement of complex platforms such as DHIS2, a widely used open source health information management system, requires significant resources, often exceeding initial budget projections. Similarly, upgrading existing digital assets to meet DPG standards can incur costs comparable to building new ones, reflecting the technical debt, as well as documentation and refactoring often required.
The DPG ecosystem, despite its collaborative ethos, is not immune to competitive pressures. Different organizations implementing similar solutions may be reluctant to adopt a shared DPG due to perceived loss of competitive advantage, differing technical approaches, or misaligned organizational priorities. This can fragment efforts and hinder the network effects that make DPGs truly powerful. Such dynamics have been observed in various open-source communities where competing versions or standards emerge, slowing down unified progress, a challenge that requires careful governance and incentive alignment to overcome, ensuring that collaboration prevails over siloed development.

Based on these experiences, we believe that the DPG ecosystem would earn substantially from the following:

\begin{enumerate}
    \item Invest selectively, but with full commitment: There is limited benefit from half-finished and insufficiently maintained solutions, but we also see that proper maintenance and documentation require resources that often exceed what many projects have available. We therefore need good mechanisms to identify promising DPGs and then support them strongly.
    \item Build robust incentives that encourage both the sharing and the adoption of DPGs. A positive example includes structuring grant conditions so that continued funding or bonus payments are tied to measurable adoption metrics, such as successful integration into a state-level platform or continued uptake by a target number of third-party organizations. This shifts the focus from mere creation to tangible impact. Detrimental practices often involve incentivizing only the initial publication of a DPG without considering its downstream usage, potentially resulting in a proliferation of DPGs that fail to achieve network effects due to lack of adoption or integration.
    \item Support national ecosystems in creating the conditions to implement DPGs in local solutions. Often there is a substantial disconnect between technical solutions developed by international actors and local realities in terms of technology used and skills available. DPG funders will need to partner with governments and other local actors to identify demand and support them in preparing to adopt DPGs developed elsewhere.
    \item Ensure continuous collaboration between creators and users of DPGs instead of only having initial consultations and final demonstrations. This also requires effort and resources, but done with a focus on building partnerships (again better fewer deeper ones), those efforts and resources will return better fit with needs and therefore uptake.
\end{enumerate}

\section{Looking ahead: How will AI-based advisory continue to look in the future?}

One recurring issue of the AIEP initiative has been that key parts of technology are constantly changing. Predicting the further path of AI is not possible, but we assume that improvements will continue, while the biggest impacts will need further time to make themselves felt as the technology is diffusing through the sector. It may be helpful to keep as a guideline that in any case the AI on which the AIEP MVPs were built will be the “worst AI we have ever used”\footnote{Adapted from \citet{Mollick2024CoIntelligence}} as the technology progresses.

Recent proprietary models such as OpenAI's o3 and GPT-5 already show substantial improvements in some tasks for certain low-resourced languages, although this does not apply to all languages, and open-source models still lack substantially behind (see \citet{Adelani2025NAACL, Adebara2025Evaluating}. The work of Digital Green, Gooey.ai, and Karya not only demonstrates how achievable progress is, but also highlights the significant obstacles that remain in bringing new technologies into production. Notably, their work raises the broader question of whether we can increasingly move away from maintaining specialized models for tasks like speech recognition, machine translation, or speech synthesis and instead allow large language models (LLMs) to handle these tasks directly. This shift could drastically improve latency by reducing the number of API calls - addressing a key current challenge.

As LLMs take on more responsibilities in the pipeline, it becomes even more crucial to improve how we evaluate where LLMs add real value and where specialized or smaller models still have the edge. Better, more nuanced evaluation methods will be critical to understanding which tasks LLMs can truly take over and which still require tailored approaches. Furthermore, contextualization will increasingly matter: solutions must take into account the nuances of the local language, the cultural context, and specific agricultural information needs. The better the underlying technology becomes, the more effectively it can leverage new agricultural content, improving the quality and relevance of advisory services.

While the optimizations discussed in the previous section remain necessary to meet user expectations, they still require significant engineering effort, frequent error fixing, and manual work—such as creating dictionaries and ontologies. The long-term goal should be to minimize these manual interventions as AI models mature, including further improvements in performance for low-resourced languages.

The cohorts are actively monitoring these advances, but uncertainty remains. For instance, it is still unclear whether a reasoning model will significantly enhance current setups, and the lack of clarity around its exact capabilities does not yet justify a full-scale reengineering. There is a consensus that advanced reasoning models will likely be useful for certain high-value tasks, but it is essential to improve identifying these tasks to justify the higher operational costs. Evidence also suggests that text-based reasoning still outperforms voice-based reasoning for many complex information tasks \citet{yang2025speechrbenchmarkspeechreasoning}.

Additionally, RAG pipelines continue to fulfill important requirements that a purely end-to-end LLM solution might not meet. RAG architectures provide a 'paper trail' that allows greater auditability and error analysis, an essential factor for many organizations.

The future of AI-driven agricultural advisory is both promising and demanding, as the remaining challenges will require collective effort to overcome. The AIEP initiative and its partners have already laid a strong foundation on which the sector can hopefully build.
Remaining challenges from better personalization and tailored advice, further integration of additional data sources and transactions, increased language coverage, and financially sustainable scale are all challenges that will become easier to overcome with collaborative effort and exchange of learnings. Lastly, with those first technological successes comes the requirement to rigorously show that these solutions can have a material positive impact on the lives of smallholder farmers. 

\section*{Acknowledgements}

This work was supported by the Gates Foundation through the AIEP project (Grant Number INV-051386) and integrated with the German Federal Ministry for Economic Cooperation and Development (BMZ) funded initiative "FAIR Forward - Artificial Intelligence for All". The conclusions and opinions expressed in this work are those of the authors alone and shall not be attributed to the Gates Foundation or the BMZ.

\bibliographystyle{plainnat}
\bibliography{references} 

\end{document}